\documentclass[preprint,showpacs,preprintnumbers,amsmath,amssymb]{revtex4}

\usepackage{graphicx}% Include figure files
\usepackage{dcolumn}% Align table columns on decimal point
\usepackage{bm}% bold math
\usepackage{float,color}

\begin{document}

\title{``Sloppy" nuclear energy density functionals (II): Finite nuclei}

\author{T. Nik\v{s}i\'{c}, M. Imbri\v{s}ak, and D. Vretenar}
\affiliation{Department of Physics, Faculty of Science, University of Zagreb, HR-10000 Zagreb, Croatia}

\date{\today}

\begin{abstract}
A study of parameter sensitivity of nuclear energy density functionals, initiated in the first part of this work \cite{NV.16}, 
is extended by the inclusion of data on ground-state properties of finite 
nuclei in the application of the manifold boundary approximation method (MBAM).  
Density functionals used in self-consistent 
mean-field calculations, and nuclear structure models based on them, are generally ``sloppy" 
and exhibit an exponential range of sensitivity to parameter variations. Concepts of information geometry 
are used to identify the presence of effective functionals of lower dimension in parameter space 
associated with parameter combinations that can be tightly constrained by data. The MBAM is used in  
an iterative procedure that systematically reduces
the complexity and the dimension of parameter space of a sloppy functional, 
with properties of nuclear matter and data on finite nuclei determining not only the values of model 
parameters, but also the optimal functional form of the density dependence.
\end{abstract}

\maketitle

%============================================================
%  Section 1
\section{\label{sec-introduction}Introduction}
%============================================================
The complexity of inter-nucleon interactions in the nuclear medium, the interplay between single-nucleon and collective degrees of freedom as well as finite size effects, make a unified treatment of the nuclear many-body problem in a single theoretical framework very difficult. Nuclear energy density functionals (EDFs), and various collective models based on them, have emerged as the most promising unified approach for a global description of nuclear structure phenomena over the entire nuclear chart. Whether nuclear EDFs are by construction microscopic, semi-empirical or fully 
phenomenological, their various implementations on the self-consistent mean-field level and collective methods that extend the 
approach beyond the mean-field approximation differ in the functional dependence on the nucleonic densities and currents, and contain model parameters. The most efficient functional density dependence and the values of most parameters, even though constrained to a certain extent by the microscopic dynamics, ultimately have to be determined by low-energy data.

In the first part of this study \cite{NV.16} we have used concepts from information geometry to analyze a representative semi-empirical functional 
and show that, in general, nuclear EDFs are ``sloppy" \cite{Transtrum.10,Transtrum.11,Gut.07,Transtrum.14,Transtrum.15}. This means 
that, even when their parameters are adjusted to data, the predictions of nuclear EDFs and related models are  
sensitive to only a few combinations of parameters ({\em stiff} parameter combinations), and exhibit an exponential decrease of sensitivity to 
variations of the remaining {\em soft} parameters that are only approximately constrained by data. 
By considering the space of model predictions as a manifold embedded in the data 
space, we have shown that the exponential distribution of model manifold widths corresponds to the range of parameter sensitivity. 
These results indicate that most nuclear EDFs, if not all, in fact contain models of lower effective dimension associated with the 
stiff combinations of model parameters. 

A systematic simplification of the complex dependence on nucleonic densities and currents of a sloppy EDF, and the 
reduction of the model to a lower dimension in parameter space, crucially depends on the selection of data that 
constrain the functional form and determine parameter values. In Ref.~\cite{NV.16} 
we have employed the Manifold Boundary Approximation Method (MBAM) \cite{Transtrum.14} to deduce the most effective 
functional form of the density-dependent coupling parameters of a representative model EDF. However, 
since the application of MBAM necessitates the computation of both first and second derivatives of observables 
with respect to model parameters along geodesic paths on the model manifold, the data used in \cite{NV.16} 
included only a set of points on a microscopic equation of state (EoS) of symmetric nuclear matter and neutron matter.  
In that case derivatives of pseudo-observables with respect to model parameters can be evaluated analytically, 
and the computational task of applying the MBAM to the nuclear system is not particularly difficult. On the downside, such a study is 
not very realistic because it does not include data on finite nuclei that are almost always used to determine or fine-tune the 
parameters of an EDF. 

In this work we extend the study of Ref.~\cite{NV.16} and use a simple numerical approximation that enables the application of 
the MBAM to realistic nuclear energy density functionals, constrained not only by the nuclear matter EoS, 
but also by observables that can be measured in nuclei all over the mass table. We will start from the 
same model EDF and microscopic EoS as in \cite{NV.16}, and include additional ground-state properties of spherical nuclei in the set of 
data used to determine the functional form and model parameters. The aim is to show how methods of information geometry, 
and the MBAM in particular, can be employed to construct and optimize nuclear energy density functionals.
Section \ref{sec-functional} defines the model functional and describes the data set used in the 
analysis of parameter sensitivity. In Sec.~\ref{sec-theoretical} we apply the MBAM in a reduction of the 
parameter space dimension and the corresponding transformation of the functional density dependence. 
 Sec.~\ref{sec-summary} contains a summary and conclusions.
%
%============================================================
%  Section 2
\section{\label{sec-functional}  The Functional DD-PC1 and the data set}
%============================================================

As representative of a class of semi-empirical energy density functional that are currently used in numerous 
studies of nuclear structure phenomena, also in this work we consider the relativistic functional DD-PC1 \cite{NVR.08}. 
It explicitly includes nucleon degrees of freedom only, and is constructed with second-order interaction terms, that is, 
the functional contains interaction terms bilinear in the densities and currents in the isoscalar-scalar, isoscalar-vector, and 
isovector-vector isospace-space channels. Many-body correlations are encoded in the 
density-dependent coupling functions: 
\begin{eqnarray}
\alpha_s(\rho)&=& a_s + (b_s + c_s x)e^{-d_s x},\nonumber\\
\alpha_v(\rho)&=& a_v +  b_v e^{-d_v x},\label{parameters} \\
\alpha_{tv}(\rho)&=& b_{tv} e^{-d_{tv} x},\nonumber
\end{eqnarray}
where the indices $s$, $v$ and $tv$ denote the isoscalar-scalar, isoscalar-vector, and 
isovector-vector channels, respectively. 
$x=\rho/\rho_\mathrm{sat}$, where $\rho_\mathrm{sat}$ indicates the nucleon
density at saturation in symmetric nuclear matter. The corresponding Lagrangian 
contains an additional derivative term with a single constant parameter \cite{NVR.08}, 
that accounts for leading effects of finite-range interactions and is essential for a 
quantitative description of nuclear density distributions. From a Lagrangian 
with bilinear interaction terms one derives the linear single-nucleon Dirac (Kohn-Sham) equation 
which, because of the density dependence of the couplings, contains also rearrangement terms. 
Like for other similar nuclear energy density functionals, both relativistic and non-relativistic, the 
explicit medium dependence of the couplings can be derived, at least in principle, from the 
underlying microscopic inter-nucleon interactions. However, the strength parameters of the functional, 
and in the present case there are ten parameters, are adjusted directly to nuclear data. 
In recent studies of global performance of relativistic EDFs in modelling ground-state properties of 
even-even nuclei over the entire mass table \cite{Agbemava.14,Agbemava.15}, it has been shown 
that DD-PC1 is currently one of the most accurate functionals, comparable in predictions to the latest 
Skyrme and Gogny non-relativistic functionals.

In this study we also consider properties of open-shell nuclei and, therefore, in addition to the effective 
interaction in the particle-hole channel, pairing correlations must be taken into account. The relativistic 
Hartree-Bogoliubov (RHB) model \cite{VALR.05} will be used in self-consistent 
mean-field calculations of ground-state properties.
As in many recent nuclear structure applications of the RHB framework based on the functional DD-PC1 \cite{NVR.11},
for the pairing interaction we employ a finite-range force that is separable in momentum space, and is completely determined 
by two parameters adjusted to reproduce the result of the D1S Gogny interaction for the density dependence of the
bell-shaped pairing gap in nuclear matter \cite{Tian.09}. 
The present analysis of parameter sensitivity of the functional DD-PC1 does not include the pairing interaction, that is, 
the parameters of the pairing force are kept constant while the functional form of the density dependence and strength 
parameters of the EDF are modified and adjusted to reproduce the data.  

Extending the analysis of Ref.~\cite{NV.16} to include data on finite nuclei, 
the (pseudo) observables that determine the parameters of the functional 
consist of two sets of data. The first contains seven points of 
the microscopic equation of state of symmetric nuclear matter, and six points of the 
neutron matter equation of state of Akmal, Pandharipande and 
Ravenhall~\cite{APR.98}, based on the Argonne $V_{18}NN$ potential and the UIX
three-nucleon interaction. One could, of course, use any other microscopic EoS of nuclear 
and neutron matter. As this study aims to demonstrate the applicability of the 
MBAM to nuclear density functionals rather than to uniquely determine the parameters of a 
functional, the choice of the microscopic EoS is not essential for the present discussion. 
In addition to the pseudo-observables of the infinite homogeneous nuclear medium, the second 
set of data contains ground-state properties of eight spherical nuclei: binding energies, 
charge radii, and available data on the difference between radii of neutron and proton distributions. 
Additional nuclei and data points could be included in a more quantitative analysis. 
Here we are interested not so much in an accurate determination of model parameters, 
but rather in qualitative constraints on the functional form of the density dependence. The 
relatively small set selected for the present analysis extends, nevertheless, from $^{16}$O to $^{214}$Pb, 
and includes both closed-shell and single open-shell nuclei. The data that will be used 
to analyze the functional form of DD-PC1 are listed in  
Tables~\ref{Tab:pseudoobservables-SNM} -- \ref{Tab:observables}.

 %----------------------------------------------------------------------------------------------------------
\begin{table}[t!]
\begin{center}
\caption{\label{Tab:pseudoobservables-SNM} Pseudo-data for infinite symmetric nuclear matter
used to compute the penalty function $\chi^2$ for the energy density functional defined 
by Eq.~(\ref{parameters}). The seven points of energy as function of the nuclear matter density 
correspond to the microscopic EoS of Akmal, 
Pandharipande and Ravenhall~\cite{APR.98}. In the least-squares fit 
the adopted error for the EoS points is 10\%.}
\bigskip
\begin{tabular}{c|c} 
\hline
\hline
{\sc pseudo-observable} &  \\ \hline \hline
$\epsilon(0.04\;fm^{-3})$        & -6.48 MeV \\
$\epsilon(0.08\;fm^{-3})$        & -12.43 MeV \\
$\epsilon(0.12\;fm^{-3})$        & -15.43 MeV \\
$\epsilon(0.16\;fm^{-3})$       & -16.03 MeV \\
$\epsilon(0.20\;fm^{-3})$       & -14.99MeV \\
$\epsilon(0.24\;fm^{-3})$       & -12.88 MeV\\
$\epsilon(0.32\;fm^{-3})$       & -6.49 MeV \\ \hline \hline
\end{tabular} 
\end{center}
\end{table}
%----------------------------------------------------------------------------------------------------------
\begin{table}[t!]
\begin{center}
\caption{\label{Tab:pseudoobservables-PNM} Pseudo-data for neutron matter
used to compute the penalty function $\chi^2$ for the energy density functional defined 
by Eq.~(\ref{parameters}). The six points correspond to the microscopic neutron
matter EoS of Akmal,  Pandharipande and Ravenhall~\cite{APR.98}. In the least-squares fit 
the adopted error for the energy of pure neutron matter as a function of density is 10\%.}
\bigskip
\begin{tabular}{c|c} 
\hline
\hline
{\sc pseudo-observable} &  \\ \hline \hline
$\epsilon(0.04\;fm^{-3})$        & 6.45 MeV \\
$\epsilon(0.08\;fm^{-3})$        & 9.65 MeV \\
$\epsilon(0.12\;fm^{-3})$        & 13.29 MeV \\
$\epsilon(0.16\;fm^{-3})$       & 17.94 MeV \\
$\epsilon(0.20\;fm^{-3})$       & 22.92 MeV \\
$\epsilon(0.24\;fm^{-3})$       & 27.49 MeV\\ \hline \hline
\end{tabular} 
\end{center}
\end{table}
%----------------------------------------------------------------------------------------------------------
\begin{table}[t!]
\begin{center}
\caption{\label{Tab:observables} The total binding energies $BE$, charge radii $r_{ch}$,
and the differences between the radii of neutron and proton density distributions
$r_{np} = r_n-r_p$ used to compute the penalty function $\chi^2$ for the energy density functional defined 
by Eq.~(\ref{parameters}). The adopted errors for the binding energies and charge radii are 
0.1$\%$ and 0.2$\%$, respectively, while $5\%$ is assumed for the accuracy of the neutron skin values.}
\bigskip
\begin{tabular}{c|cccc} 
\hline
\hline
{\sc Nucleus} & $BE$ (MeV) & $r_{ch}$ (fm) & $r_n-r_p$ (fm)   \\ \hline \hline
$^{16}$O        & -127.619 MeV & 2.73& \\
%$^{36}$S        & -308.714 MeV & 3.299&\\
$^{48}$Ca      &-415.991 MeV & 3.484& \\
$^{72}$Ni       & -613.173 MeV &   & \\
$^{90}$Zr       & -783.893 MeV & 4.272\\
$^{116}$Sn     & -988.681 MeV & 4.626 & 0.12\\ 
%$^{124}$Sn     & -1049.962 MeV&  4.674 & 0.19\\ 
$^{132}$Sn     & -1102.860 MeV &  &  & \\ 
%$^{204}$Pb     & -1607.446 MeV & 5.486 & \\ 
$^{208}$Pb     & -1636.446 MeV & 5.505 &0.20\\ 
$^{214}$Pb     & -1663.298 MeV & 5.562& \\ 
%$^{210}$Po     & -1645.228 MeV & && \\  \hline \hline
\end{tabular} 
\end{center}
\end{table}
%----------------------------------------------------------------------------------------------------------
%----------------------------------------------------------------------------------------------------------
\begin{figure}[htb]
\centering
\includegraphics[scale=0.45]{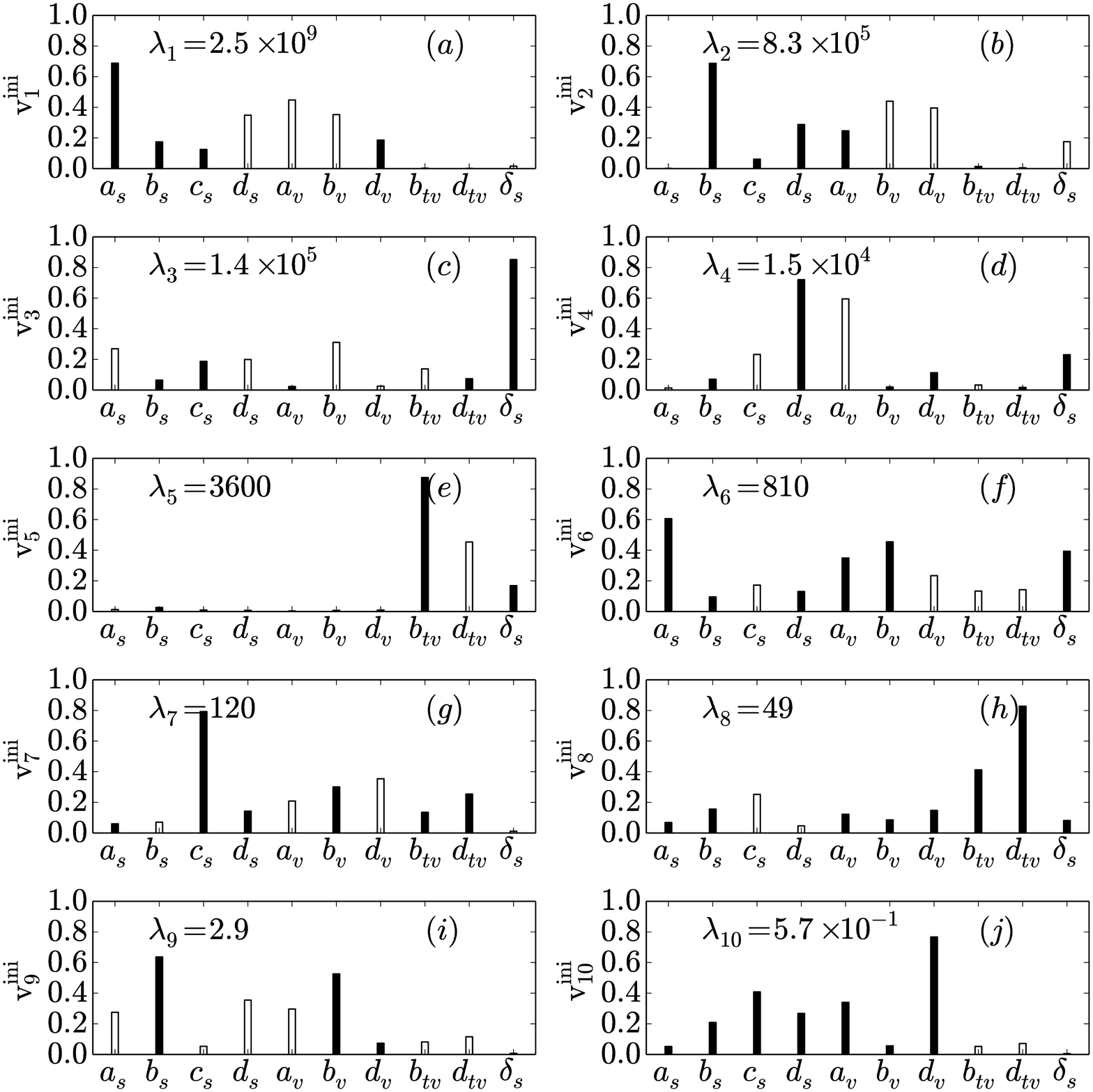} 
\begin {center}
\caption{\label{fig:red1_eigenvectors} Eigenvectors and eigenvalues of the
$10\times 10$ Hessian matrix of second derivatives $\mathcal{M}$ of $\chi^2(\mathbf{p})$ 
at the best-fit point for the functional defined by the couplings of Eq.~(\ref{parameters}), 
plus the strength parameter of the derivative term $\delta_s$. 
The empty and filled bars indicate that the corresponding amplitudes contribute with 
opposite signs. }
\end{center}
\end{figure}
%----------------------------------------------------------------------------------------------------------

The set of all possible values of model parameters defines the 10-dimensional manifold 
embedded in the $N$-dimensional data space ($N = 29$ in the present case). 
For a given point in the data space, 
that is, for the set of data listed in Tables~\ref{Tab:pseudoobservables-SNM} -- \ref{Tab:observables},
the nine parameters that determine the density dependence of the 
coupling functions Eq.~(\ref{parameters}), plus the strength parameter of the derivative term 
$\delta_s (\partial_\nu \bar{\psi}\psi)  (\partial^\nu \bar{\psi}\psi)$ of the functional DD-PC1, 
are optimized by minimizing the penalty function
$\chi^2(\mathbf{p})$ on the manifold of model predictions embedded in the data space:
\begin{equation}
  \chi^2(\mathbf{p}) 
  = \sum_{n=1}^N{ r_n^2(\mathbf{p}) },
\end{equation}
where $r_n(\mathbf{p})$ denotes the residual
\begin{equation}
r_n(\mathbf{p}) = \frac{\mathcal{O}_n^{(mod)}(\mathbf{p})-\mathcal{O}_n}
                         {\Delta \mathcal{O}_n} ,
\end{equation}
and $\mathcal{O}_n^{(mod)}$ are model predictions that depend on the set of parameters $\mathbf{p}=\{p_1,\dots,p_F\}$. 
Every observable is weighted by the inverse of $\Delta \mathcal{O}_n$, and the adopted errors are given in the captions to 
Tables~\ref{Tab:pseudoobservables-SNM} -- \ref{Tab:observables}. The behavior of the model around the best-fit point 
$\mathbf{p}_0$ can be analyzed in the quadratic approximation to the penalty function: 
\begin{equation}
  \Delta \chi^2(\mathbf{p}) = 
  \chi^2(\mathbf{p}) - \chi^2(\mathbf{p}_0)  =
  \frac{1}{2} \Delta \mathbf{p}^T \hat{\mathcal{M}} \Delta \mathbf{p}\;, 
\end{equation} 
 where $\Delta \mathbf{p}    = \mathbf{p} - \mathbf{p}_0$. The  curvature matrix
\begin{equation}
  \mathcal{M}_{\mu \nu}  = 
   \left. \frac{\partial^2 \chi^2 }{\partial p_\mu \partial p_\nu} \right|_{\mathbf{p}=\mathbf{p}_0},
      \label{M} 
\end{equation}
is symmetric and can be
diagonalized by an orthogonal transformation: 
$\hat{\mathcal{M}}=\hat{\mathcal{A}}\hat{\mathcal{D}}\hat{\mathcal{A}}^T$,
where $\hat{\mathcal{A}}$ denotes the orthogonal matrix with columns
corresponding to normalized eigenvectors of $\hat{\mathcal{M}}$, and 
the diagonal matrix $\hat{\mathcal{D}}$ contains the 
eigenvalues of $\hat{\mathcal{M}}$.  The deviation of 
$\chi^2$ from its minimum value can be expressed as 
\begin{equation}
  \Delta \chi^2(\mathbf{p}) 
  = \frac{1}{2}
  \Delta \mathbf{p}^T \left( \mathcal{A}\mathcal{D}\mathcal{A}^T  \right)
   \Delta \mathbf{p} 
  = \frac{1}{2}
  \mathbf{\xi}^T \mathcal{D} \mathbf{\xi} = \frac{1}{2} \sum_{\alpha =1}^F{\lambda_\alpha
    \xi_\alpha^2} .
\label{delta_chi}
\end{equation}
%------------------
The transformed vectors $\mathbf{\xi}=\hat{\mathcal{A}}^T\mathbf{p}$ define 
the principal axes on the $F$-dimensional model manifold. 
Fig.~\ref{fig:red1_eigenvectors} displays the eigenvectors and eigenvalues of the
$10\times 10$ Hessian matrix of second derivatives $\mathcal{M}$ of $\chi^2(\mathbf{p})$ 
at the best-fit point for the functional defined by the couplings of Eq.~(\ref{parameters}), 
plus the strength parameter of the derivative term $\delta_s$. We will refer to this set of parameters as {\sc SET 1}.
The empty and filled bars indicate that the corresponding amplitudes contribute with 
opposite signs. The eigenvalues of the Hessian matrix, characterizing the sensitivity of model predictions 
to variations along orthogonal directions in parameter space, span ten orders of magnitude, and this is 
typical of sloppy models that exhibit an exponentially large range of sensitivities to changes in 
parameter values \cite{BS.03,Wat.06,Gut.07,Transtrum.10,Machta.13,Transtrum.15}. Such a model 
is essentially determined by only a  few {\em stiff} directions in parameter space 
characterized by large eigenvalues $\lambda_\alpha$, while the remaining 
{\em soft} directions that correspond to small eigenvalues $\lambda_\alpha$ are not 
constrained by the data used in the least-squares fit. 

%============================================================
\section{\label{sec-theoretical}Model reduction by the Manifold Boundary Approximation Method }
%============================================================

In Ref.~\cite{NV.16} we have also shown that the functional DD-PC1 exhibits another 
unique characteristic of sloppy models \cite{Transtrum.10,Transtrum.11,Transtrum.14,Transtrum.15}, 
namely that the widths of the model manifold in the directions of the eigenvectors of the 
Hessian follow closely the distribution of sensitivity (square root of the eigenvalue of the Hessian) 
of the functional to changes in the values of the corresponding parameter combinations.
By interpreting the space of model predictions as a manifold embedded in the Euclidean data space, with 
parameters of the functional as coordinates on the manifold, one can explore the boundaries of the manifold
using geodesic paths. Boundaries correspond to points on the 
manifold where the metric becomes singular, and the arc length of geodesics, along directions specified 
by the eigenvectors of the Hessian matrix at the minimum of $\chi^2$, provide a measure of the manifold 
width in each of these directions \cite{Transtrum.10}.
The parameters corresponding to a geodesic path can be found as the solution of the 
second-order differential equation
\begin{equation}
\label{eq:geodesic-equation}
\ddot{p}_\mu + 
\sum_{\alpha \beta} {\Gamma_{\alpha \beta}^\mu \dot{p}_\alpha \dot{p}_\beta } =0,
\end{equation}
where $\Gamma_{\alpha \beta}^\mu$ are the connection coefficients: 
\begin{equation}
\label{eq:connection-coefficients}
\Gamma_{\mu \nu}^\alpha = \sum_{\beta}{(g^{-1})_{\alpha \beta}
\sum_m{ \frac{\partial r_m}{\partial p_\beta}  \frac{\partial^2 r_m}{\partial p_\mu \partial p_\nu}}} \;,
\end{equation}
the metric on the model manifold is defined by the Fisher information matrix (FIM):
\begin{equation}
\label{eq:metric-tensor}
{g}_{\mu \nu} =  \sum_m{\frac{\partial r_m}{\partial p_\mu}\frac{\partial r_m}{\partial p_\nu}} \;,
\end{equation}
and the dot denotes differentiation with respect
to the affine parametrisation of the geodesic. Note that at the best-fit point (minimum of $\chi^2$),
the metric ${g}_{\mu \nu}$ of the model manifold approximately equals the Hessian matrix of 
second derivatives $\mathcal{M}$ of $\chi^2(\mathbf{p})$. 
The geodesic equation presents an 
initial value problem in the parameter space. Starting from any point on the model 
manifold, one follows the geodesic path in a given direction until the boundary 
is identified by the metric tensor becoming singular. In particular, if one considers the 
best-fit point $\chi^2(\mathbf{p}_0)$, the geodesic equation can be integrated along the 
eigendirection of the Hessian matrix to determine the corresponding 
boundaries of the model manifold. The initial value $\mathbf{p}_{ini}$ corresponds 
to the best-fit parameters, and the initial velocities
$\dot{\mathbf{p}}_{ini}$ are determined by the eigenvectors of the Hessian 
at the best-fit point. An eigenvector defines two possible
directions for integration (positive and negative),  and the sum of the two 
arc lengths equals the width of the manifold for that particular eigendirection~\cite{Transtrum.14}.

In the analysis of the DD-PC1 model manifold of Ref.~\cite{NV.16} we have only considered a set of pseudo-observables, 
energies as function of density, for infinite nuclear matter. In this case the derivatives of residuals with 
respect to model parameters, contained in the expression for the connection coefficients (\ref{eq:connection-coefficients}), 
can be calculated analytically. Here the data set is extended with ground-state properties of finite nuclei for which the 
connection coefficients have to be calculated numerically. The computational task can be considerably 
reduced by interchanging the order of summations implicit in Eqs.~(\ref{eq:geodesic-equation}) and (\ref{eq:connection-coefficients}), that is,
by calculating first the following quantity:
\begin{equation}
\sum_{\alpha \beta}{\frac{\partial^2 r_m}{\partial p_\alpha \partial p_\beta} \dot{p}_\alpha \dot{p}_\beta}
 = \lVert \dot{\mathbf{p}} \rVert^2
 \sum_{\alpha \beta}{\frac{\partial^2 r_m}{\partial p_\alpha \partial p_\beta} 
 \frac{\dot{p}_\alpha}{\lVert \dot{\mathbf{p}} \rVert}  \frac{\dot{p}_\beta}{\lVert \dot{\mathbf{p}} \rVert}}
=  \frac{\lVert \dot{\mathbf{p}} \rVert^2}{\epsilon^2}
 \sum_{\alpha \beta}{\frac{\partial^2 r_m}{\partial p_\alpha \partial p_\beta} 
  \delta p_\alpha \delta p_\beta }\;,
\end{equation}
where 
\begin{equation}
\delta p_\mu \equiv \epsilon \frac{\dot{p}_\mu }{\lVert \dot{\mathbf{p}} \rVert} ,\quad \mu \in \{\alpha,\beta\} \;,
\end{equation}
and $\epsilon$ is a small constant. A Taylor 
expansion for the residual $r_m$ leads to the following expression:
\begin{align}
 \sum_{\alpha \beta}{\frac{\partial^2 r_m}{\partial p_\alpha \partial p_\beta} 
  \delta p_\alpha \delta p_\beta } &\approx r_m (p_1+\delta p_1,\dots,p_n+\delta p_n)
  + r_m (p_1-\delta p_1,\dots,p_n-\delta p_n) \nonumber \\
  &-2r_m (p_1,\dots,p_n).
\end{align}
The calculation is simplified by numerically computing directional second order derivatives instead of all second derivatives entering
the definition of connection coefficient. The initial nine parameters defined in Eq.~(\ref{parameters}), plus the strength parameter of 
the derivative term $\delta_s$, are transformed as follows:
\begin{equation}
a_s = a_{s,bf} p_{a_s} , \quad b_s = b_{s,bf} p_{b_s}, \quad c_s = c_{s,bf} p_{c_s} ,\quad
d_s = d_{s,bf} p_{d_s} ,
\end{equation}
\begin{equation}
a_v = a_{v,bf} p_{a_v} , \quad b_v = b_{v,bf} p_{b_v}, \quad d_v = d_{v,bf} p_{d_v} ,
\end{equation}
\begin{equation}
b_{tv} = b_{tv,bf} p_{b_{tv}}, \quad d_{tv} = d_{tv,bf} p_{d_{tv}} ,
\end{equation}
\begin{equation}
\delta_{s} = \delta_{s,bf} p_{\delta_{s}}.
\end{equation}
where the subscript $bf$ denotes the best-fit values obtained by minimizing the penalty function
$\chi^2(\mathbf{p})$ ({\sc SET 1}). In this way all the parameters in the geodesic equation (\ref{eq:geodesic-equation}) 
become dimensionless,  and their values at the initial point:
\begin{equation}
p_\mu(0) =1, \quad \mu \in \{a_s,b_s,c_s,d_s,a_v,b_v,d_v,b_{tv},d_{tv},\delta_s\} .
\end{equation}
Compared to the transformation we used in the previous study (see Appendix B of Ref.~\cite{NV.16}),
here the parameters are not constrained to have the same sign along the geodesic path. 
This allows us to explore the entire parameter space and, by including data on finite nuclei, the data 
space contains enough points to avoid possible unphysical regions of parameters. 
The initial velocities are determined by the corresponding amplitudes of  
eigenvectors of the Hessian matrix 
($\mathcal{M} = \mathcal{A} \mathcal{D} \mathcal{A}^T$)
of the penalty function (cf. Fig.~\ref{fig:red1_eigenvectors}):
\begin{equation}
\label{eq:direction1}
\dot{p}_\mu(0) \sim \mathcal{A}_{\mu}.
\end{equation}
The overall normalization factor is chosen so that the data space norm of the
velocity vector equals one: 
\begin{equation}
\label{eq:direction2}
\sum_{\mu,\nu}{g_{\mu \nu} \dot{p}_\mu(0)  \dot{p}_\nu(0)} =1,
\end{equation}
and $g_{\mu \nu}$ denotes the metric tensor (FIM). 
Because an eigenvector is defined up to an 
overall phase, Eqs.~(\ref{eq:direction1}) and (\ref{eq:direction2}) determine 
two opposite directions for the initial velocity. For each direction 
the geodesic equation is integrated up to the manifold boundary and, 
since the data space norm of the velocity remains constant, the length of the traversed 
path in the data space equals the maximal value of the affine parameter. The sum of the 
two arcs equals the width of the model manifold for this particular combination of bare 
model parameters. 

The resulting widths of the model manifold in the directions of eigenvectors 
of the Hessian matrix of the penalty function $\chi^2(\mathbf{p}_0)$ exhibit an exponential 
distribution of values (cf. Fig. 1 in Ref.~\cite{NV.16}), 
and this points to the existence of an effective functional of lower 
dimension associated with stiff parameter combinations. The reduction of a general sloppy 
model to lower dimension in parameter space is essentially determined by the choice of data 
to which the parameters are adjusted. Following our study of DD-PC1 in nuclear matter 
in Ref.~\cite{NV.16}, we employ the Manifold Boundary Approximation Method (MBAM) 
\cite{Transtrum.14} to construct a simpler EDF of lower parameter space dimension, 
constrained by the data set listed in Tables~\ref{Tab:pseudoobservables-SNM}-\ref{Tab:observables}, 
and which also includes ground-state data of finite nuclei.  

Starting from the best-fit point $\mathbf{p}_0$ in parameter space, the geodesic equation is 
integrated in the eigendirection that corresponds to the smallest eigenvalue of the Hessian matrix, 
until the boundary of the model manifold is reached. Because the eigenvector is defined up to an 
overall phase, we choose the direction in which the parameter space norm of
the velocity vector ($\sum_\mu{\dot{p}_\mu^2}$) increases \cite{NV.16,Transtrum.14}.
The model limit associated with the manifold boundary 
is analyzed and a new model is constructed that contains one less parameter. 
The new model is then optimized by a least-squares fit to the same set of data, and used as a starting point 
for the next iteration of the MBAM. This method, therefore, reduces the sloppiness of a model by 
successively eliminating soft combinations of bare parameters. In the ideal case in which the data set contains 
all the information necessary to completely determine an {\it a priori} unknown physical model, the MBAM 
will produce a unique non-sloppy model with only stiff combinations of bare parameters. 
%----------------------------------------------------------------------------------------------------------
\begin{figure}[htb]
\centering
\includegraphics[scale=0.6]{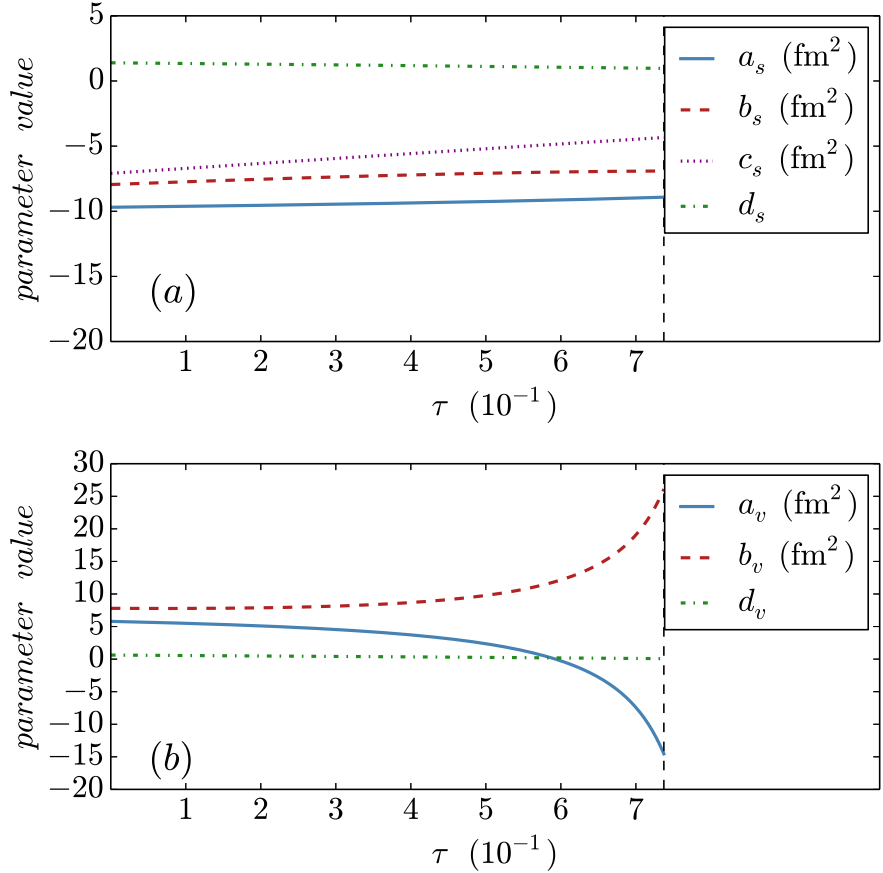} 
\begin {center}
\caption{\label{fig:parameters1-step1} (Color online) The 
parameters of the isoscalar part of the ten-parameter functional {\sc SET 1},
as functions of the affine parametrisation, along the geodesic path determined 
by the eigenvector of the Hessian matrix that corresponds to the smallest eigenvalue (cf. Fig.~\ref{fig:red1_eigenvectors}). 
The panel (a) displays the four parameters of the scalar channel, while the evolution of the three 
parameters that determine the vector channel is shown in the panel (b). 
}
\end{center}
\end{figure}
%----------------------------------------------------------------------------------------------------------
\begin{figure}[htb]
\centering
\includegraphics[scale=0.6]{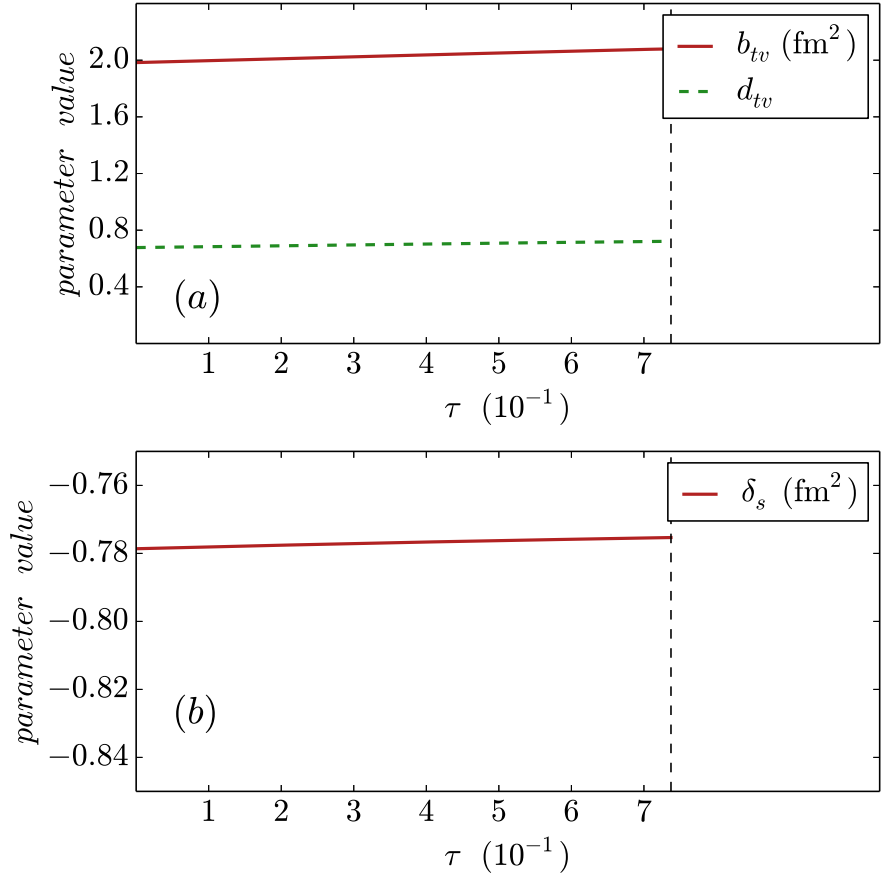} 
\begin {center}
\caption{\label{fig:parameters2-step1} (Color online) Same as in the caption to Fig.~\ref{fig:parameters1-step1} 
but for the two parameters of the isovector part of the functional defined in 
Eq.~(\ref{parameters}) (panel (a)), and the strength parameter of the derivative term (panel (b)).}
\end{center}
\end{figure}
%----------------------------------------------------------------------------------------------------------
\begin{figure}[htb]
\centering
\includegraphics[scale=0.5]{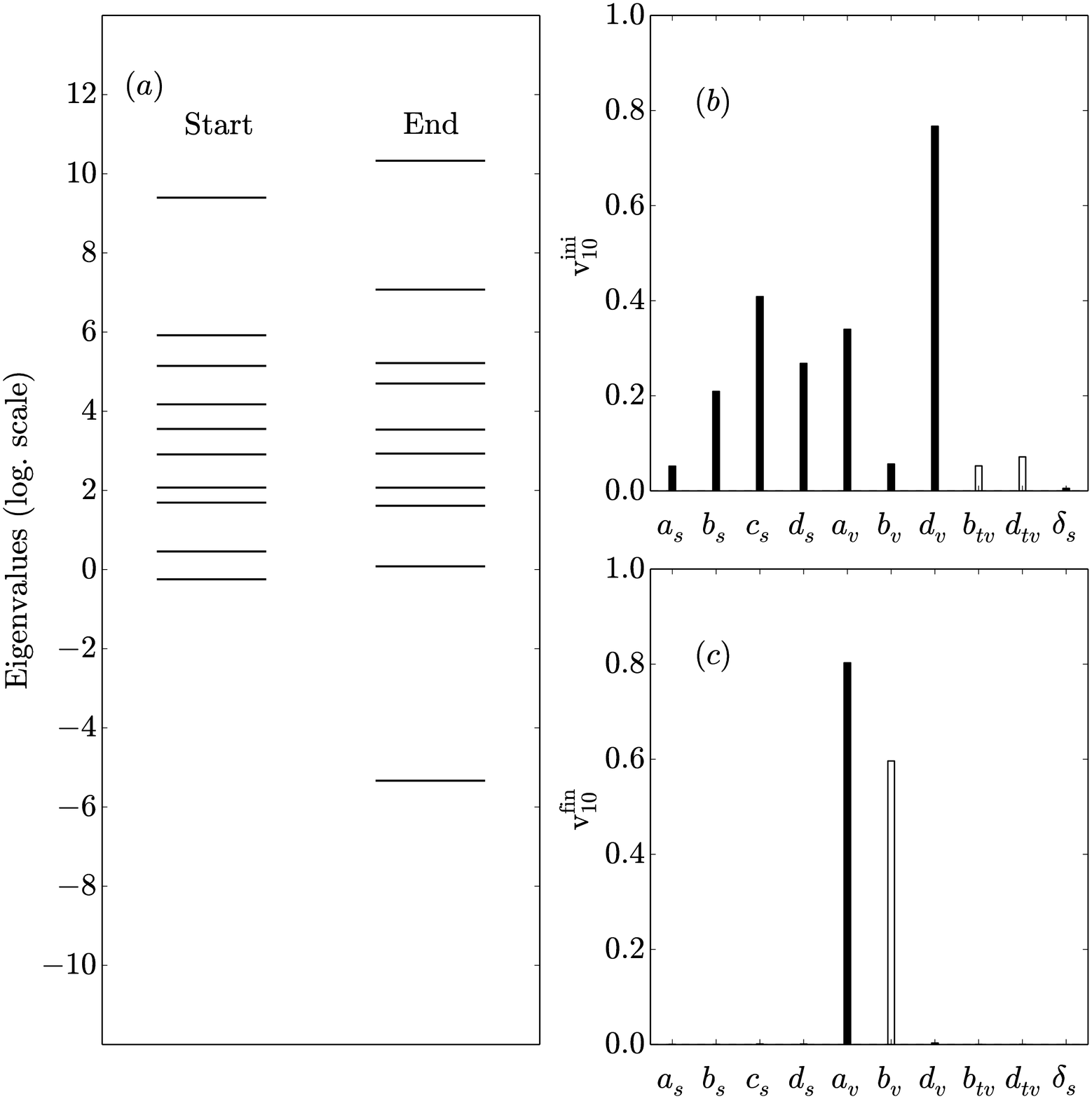} 
\begin {center}
\caption{\label{fig:spectrum-step1} The initial (best-fit point) 
and final (at the boundary of the model manifold) eigenspectrum of the FIM 
for the ten-parameter functional {\sc SET 1} (panel (a)).  The 
eigenvectors that correspond to the initial and final smallest eigenvalues 
are shown in panels (b) and (c). 
}
\end{center}
\end{figure}
%----------------------------------------------------------------------------------------------------------

%%%%%%%%%        First iteration         %%%%%%%%%%%%%%%%%%%%%%%%%
The first iteration of the MBAM for the initial ten-parameter functional is illustrated in Figs.~\ref{fig:parameters1-step1} -- 
\ref{fig:spectrum-step1}. The evolution of the parameters along the geodesic path determined 
by the eigenvector of the Hessian matrix that corresponds to the smallest eigenvalue,  
as functions of the affine parametrisation of the geodesic, is shown in Figs.~\ref{fig:parameters1-step1}
and~\ref{fig:parameters2-step1}. The geodesic equation is integrated with the initial conditions
described above until the corresponding boundary of the model manifold is identified. While there are no significant changes 
of the parameters in the isoscalar-scalar and isovector-vector channels, as well as the strength of the 
derivative term, in the isoscalar-vector channel one notes that the parameters 
 $a_v$ and $b_v$ start to diverge as the geodesic path approaches the boundary of the manifold, and $d_v$ 
tends to a small value close to zero. 
This can be understood from the form of the vector coupling:
\begin{equation}
\alpha_v(\rho) = a_v + b_v e^{-d_v x}, \quad  x=\rho/\rho_{sat}.
\end{equation}
When $d_v$ approaches zero, the derivatives 
$\partial \alpha_v/\partial a_v$ and $\partial \alpha_v/\partial b_v$ are virtually identical, 
the corresponding rows/columns of the FIM are almost equal and the matrix
becomes singular. In Fig.~\ref{fig:spectrum-step1} we plot 
the initial and final (at the boundary) eigenspectrum of the FIM in the panel (a), and the initial and final eigenvectors 
that correspond to the smallest eigenvalues in panels (b) and (c). At the boundary the smallest 
eigenvalue separates from the rest of the spectrum and tends to zero.  While all bare parameters, 
with the exception of $\delta_s$, contribute to the amplitudes of the initial 
softest eigenvector, at the boundary only the components $a_v$ and $b_v$ (with opposite phases) 
determine the eigenvector of the FIM with the eigenvalue  approaching zero. 
The limiting behavior of $a_v$, $b_v$ and $d_v$ suggests 
the following Taylor expansion for the vector coupling function at the boundary: 
\begin{equation}
\label{eq:9_parameters}
\alpha_v(\rho) \approx a_v + b_v(1-d_vx) = a_v + b_v -b_vd_v x = \tilde{a}_v + \tilde{b}_v x .
\end{equation}
In first order this reduces the functional form of the coupling function in the isoscalar-vector channel to a linear density dependence, 
and the corresponding number of bare parameters from three to two. In the final step the nine parameters of the resulting model 
are again determined in a least-squares fit to the data listed Tables~\ref{Tab:pseudoobservables-SNM}-\ref{Tab:observables}
({\sc SET 2}), and used as a starting point for the next iteration of the MBAM.

%%%%%%%%%%%       Second iteration      %%%%%%%%%%%%%%%%%%%%%%%%%%

%----------------------------------------------------------------------------------------------------------
\begin{figure}[htb]
\centering
\includegraphics[scale=0.45]{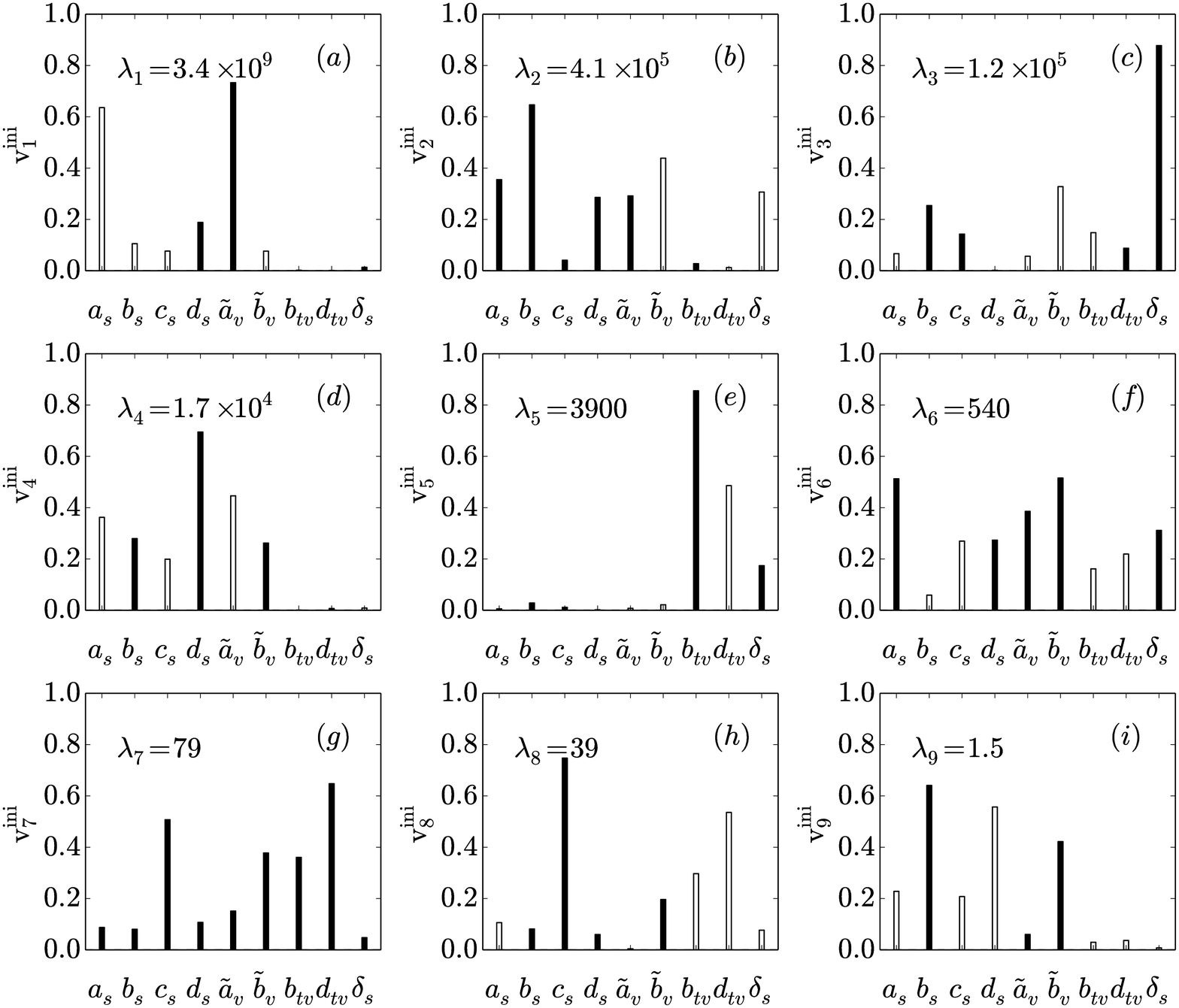} 
\begin {center}
\caption{\label{fig:red2_eigenvectors} Same as in the caption to Fig.~\ref{fig:red1_eigenvectors} 
but for the functional with nine parameters ({\sc SET 2}) obtained by applying the MBAM to the ten-parameter 
functional {\sc SET 1}.
}
\end{center}
\end{figure}
\begin{figure}[htb]
\centering
\includegraphics[scale=0.6]{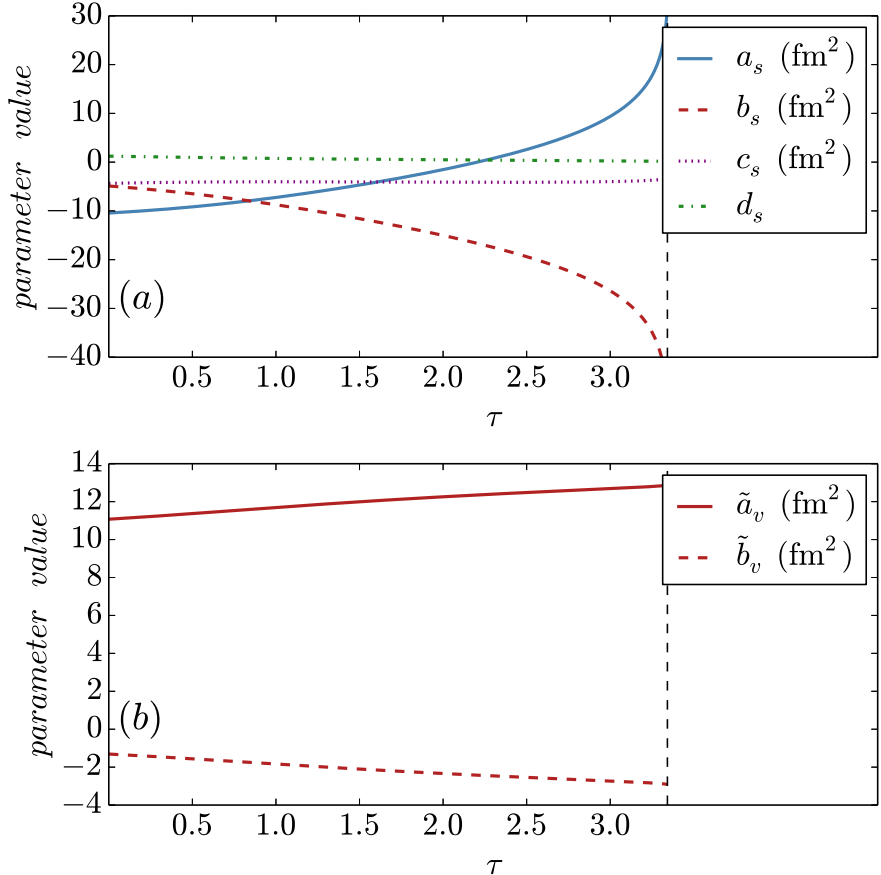} 
\begin {center}
\caption{\label{fig:parameters1-step2} (Color online) 
Same as in the caption to Fig.~\ref{fig:parameters1-step1} but for the nine-parameter functional ({\sc SET 2}).
}
\end{center}
\end{figure}
\begin{figure}[htb]
\centering
\includegraphics[scale=0.6]{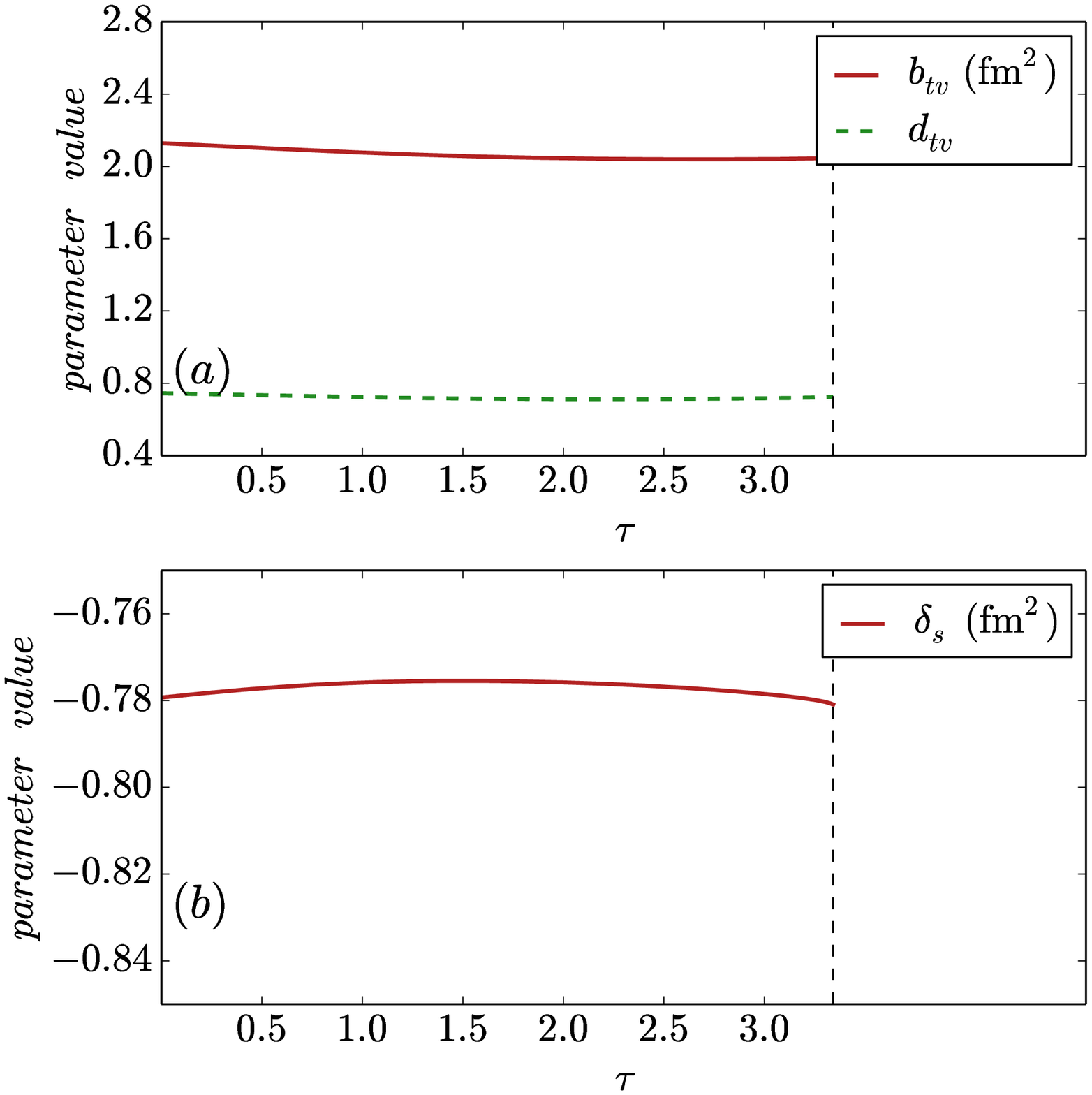} 
\begin {center}
\caption{\label{fig:parameters2-step2} (Color online) 
Same as in the caption to Fig.~\ref{fig:parameters2-step1} but for the nine-parameter functional ({\sc SET 2}).
}
\end{center}
\end{figure}
\begin{figure}[htb]
\centering
\includegraphics[scale=0.5]{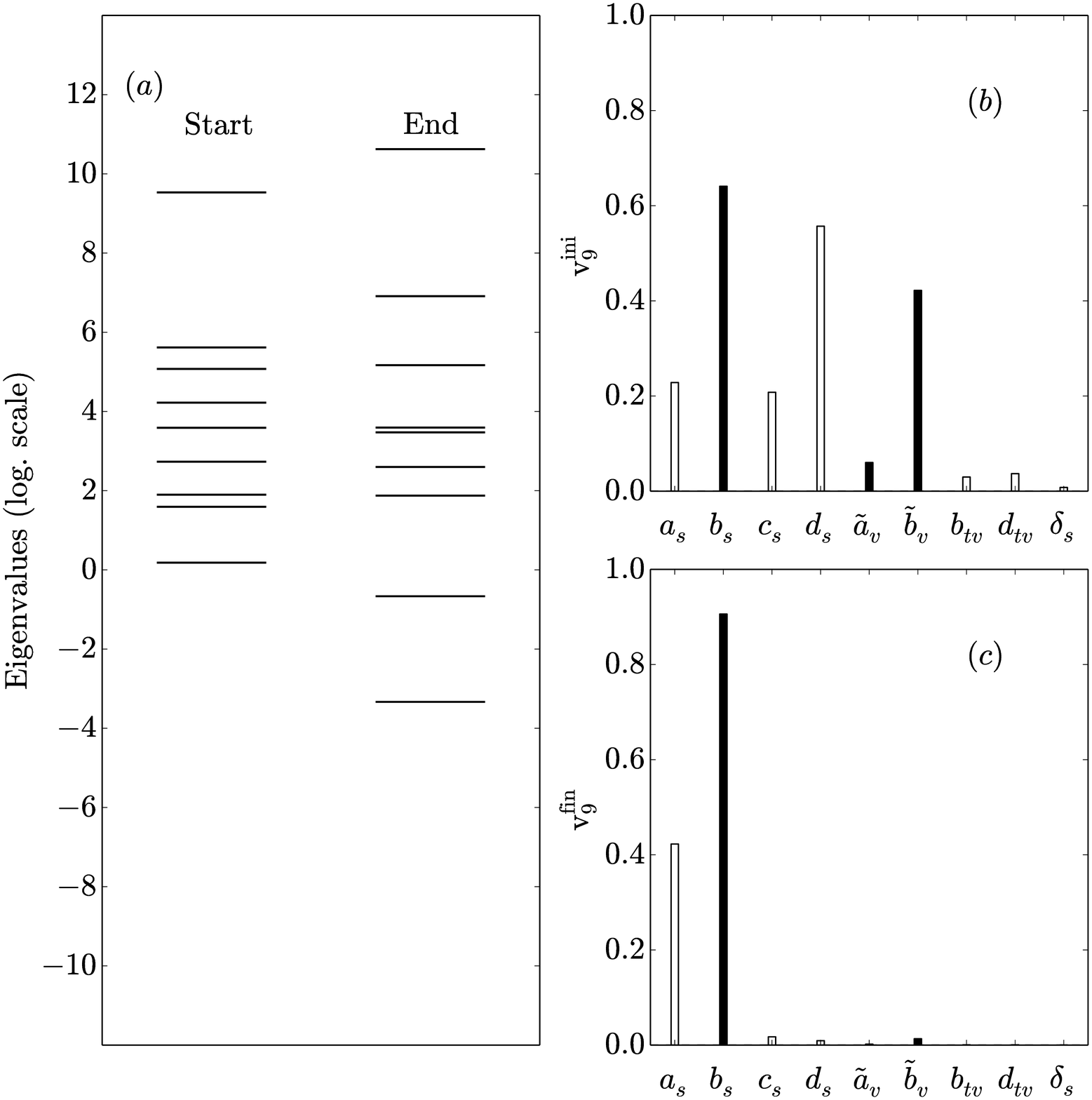} 
\begin {center}
\caption{\label{fig:spectrum-step2} 
Same as in the caption to Fig.~\ref{fig:spectrum-step1} but for the nine-parameter functional ({\sc SET 2}).
}
\end{center}
\end{figure}

%----------------------------------------------------------------------------------------------------------
The second iteration thus starts with a model functional determined by the nine parameters:
$a_s$, $b_s$, $c_s$, $d_s$, $\tilde{a}_v$, $\tilde{b}_v$, $b_{tv}$, $d_{tv}$ and $\delta_s$ ({\sc SET 2}).
The eigenvectors and eigenvalues of the FIM calculated at the best-fit point (Hessian matrix) 
are shown in Fig.~\ref{fig:red2_eigenvectors}. In this case the eigenvalues span nine orders of magnitude. 
Starting from the best-fit point, the geodesic equation is integrated following the
direction determined by the softest eigenvector, and in Figs.~\ref{fig:parameters1-step2} and~\ref{fig:parameters2-step2} 
we display the evolution of the model parameters, as functions of the affine parametrisation of the geodesic. 
A scenario similar to the first iteration unfolds, only this time in the isoscalar-scalar channel. 
As the parameter of the exponential function $d_s$ approaches zero, and the two parameters $a_s$ and $b_s$ 
start to diverge, the FIM becomes singular at the manifold boundary. 
The eigenvector with the smallest eigenvalue decouples from the rest, as shown 
in Fig.~\ref{fig:spectrum-step2}. The eigenvalue tends to zero, 
while the amplitudes of the corresponding eigenvector exhibit dominant out-of-phase components
 $a_s$ and $b_s$. The initial and final eigenvalues of the FIM are displayed in
the panel (a) of Fig.~\ref{fig:spectrum-step2}. 

The divergent behavior of the parameters $a_s$ and $b_s$, and the 
Taylor expansion of the exponential coupling function to first order in the small parameter 
$d_s$ at the boundary, lead to the following reduction of the isoscalar-scalar coupling function:
\begin{equation}
\label{eq:8_parameters}
\alpha_s(\rho) \approx a_s + (b_s+c_s x)(1-d_sx) = a_s + b_s + (c_s -b_sd_s) x - c_s d_s x^2 
= \tilde{a}_s + \tilde{b}_s x + \tilde{c}_s x^2 .
\end{equation}
The second iteration, therefore, transforms the coupling in the isoscalar-scalar channel to a polynomial 
of second degree in the nucleonic density, and the number of parameters is reduced by one. As a final 
step of this MBAM iteration, the new model determined by the 
eight parameters: $\tilde{a}_s$, $\tilde{b}_s$, $\tilde{c}_s$,
$\tilde{a}_v$, $\tilde{b}_v$, $b_{tv}$, $d_{tv}$ and $\delta_s$, is again fitted to the 
observables listed in Tables~\ref{Tab:pseudoobservables-SNM}-\ref{Tab:observables}. The 
resulting parameters are denoted {\sc SET 3}. 

Note that in both MBAM iterations the parameters of the isovector-vector channel $b_{tv}$ and $d_{tv}$, as well as 
the strength parameter of the derivative term $\delta_s$, do not display significant variations along the softest eigendirections 
(Figs.~\ref{fig:parameters2-step1} and \ref{fig:parameters2-step2}), or in the two successive least-squares adjustments to data. 
This means that the parameters of the isovector channel and the derivative term are already constrained by the data used in 
the fit (in particular, the neutron matter EoS and difference of the radii of neutron and proton distributions for the isovector channel, and 
charge radii for the derivative term). The dominant component of the third eigenvector of the Hessian matrix for both 
best-fit points (Figs.~\ref{fig:red1_eigenvectors} and \ref{fig:red2_eigenvectors}) corresponds to the bare parameter $\delta_s$, 
while the fifth eigenvector is characterized by large out-of-phase amplitudes that correspond to $b_{tv}$ and $d_{tv}$. To 
constrain the in-phase combination of the isovector parameters additional data are required such as, for instance, 
information on the isovector effective mass in nuclear matter. 
%----------------------------------------------------------------------------------------------------------
%----------------------------------------------------------------------------------------------------------
%----------------------------------------------------------------------------------------------------------
%----------------------------------------------------------------------------------------------------------

%----------------------------------------------------------------------------------------------------------
%----------------------------------------------------------------------------------------------------------
The functional determined by the eight parameters: five in the isoscalar channel:
\begin{equation}
\alpha_s(\rho) = \tilde{a}_s + \tilde{b}_s x + \tilde{c}_s x^2 \quad \textnormal{and} \quad
\alpha_v(\rho) = \tilde{a}_v + \tilde{b}_v x,
\end{equation}
where $x = \rho/\rho_{sat}$, two in the isovector-vector channel $b_{tv}$ and $d_{tv}$, and the
parameter of the derivative term $\delta_s$ ({\sc SET 3}), could, in principle, be further reduced. However, for the 
data set of Tables~\ref{Tab:pseudoobservables-SNM}-\ref{Tab:observables}, in the next, third iteration, the 
integration of the geodesic equation does not lead to the decoupling of the softest eigenvector. Even considering 
both directions determined by the softest eigenvector of the Hessian matrix, we have not been able 
to reach the corresponding boundary of the model manifold and, therefore, the number of parameters could not be reduced. 
Note that in our previous study \cite{NV.16}, in which only nuclear matter pseudo-data were used to 
determine the parameters of the functional, it was possible to reduce both isoscalar coupling functions to a 
linear dependence of the nucleonic density, that is, a third iteration reduced the number of isoscalar parameters to 
four. Here this is no longer possible because ground-state properties of finite nuclei are included in the data set and, 
with only a simple linear density dependence of the scalar and vector coupling functions, a self-consistent mean-field 
calculation based on such a functional could not reproduce the data. 

%----------------------------------------------------------------------------------------------------------
\begin{figure}[htb]
\centering
\includegraphics[scale=0.55]{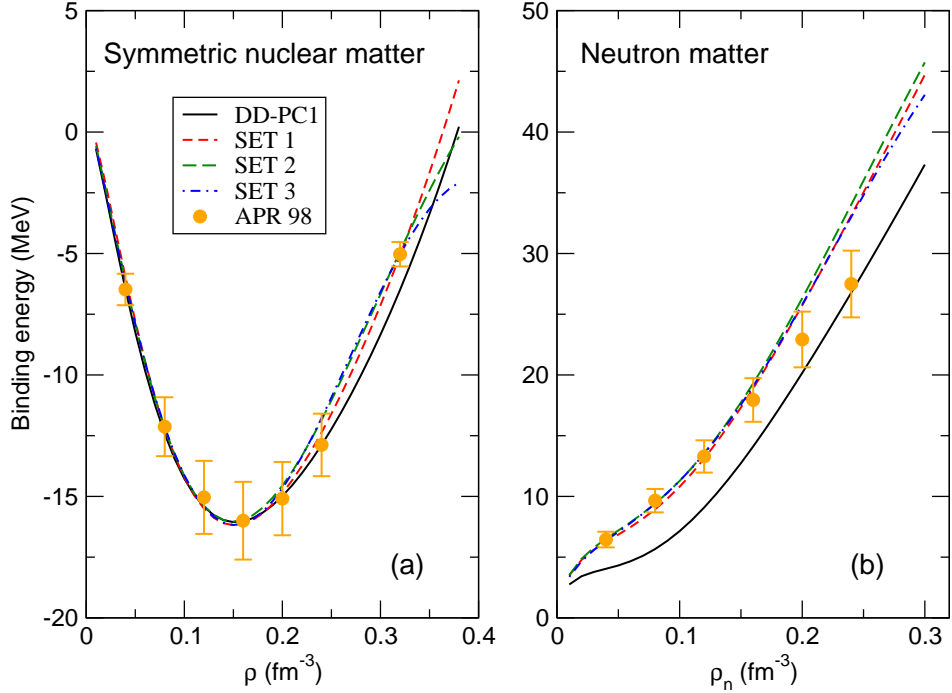} 
\begin {center}
\caption{\label{fig:EoS-SNM-PNM} (Color online) 
Equation of state of symmetric nuclear matter (lpanel (a)) and neutron matter (panel (b)).
Energy as a function of nuclear (neutron) matter calculated with the functional DD-PC1, and 
the three functionals adjusted in this study {\sc SET 1}, {\sc SET 2}, and {\sc SET3}, are 
shown in comparison to the microscopic equations of state~\cite{APR.98}. 
The points to which the parameters of 
the functionals have been fitted are shown with the adopted 
uncertainty 10$\%$.
}
\end{center}
\end{figure}
\begin{figure}[htb]
\centering
\includegraphics[scale=0.55]{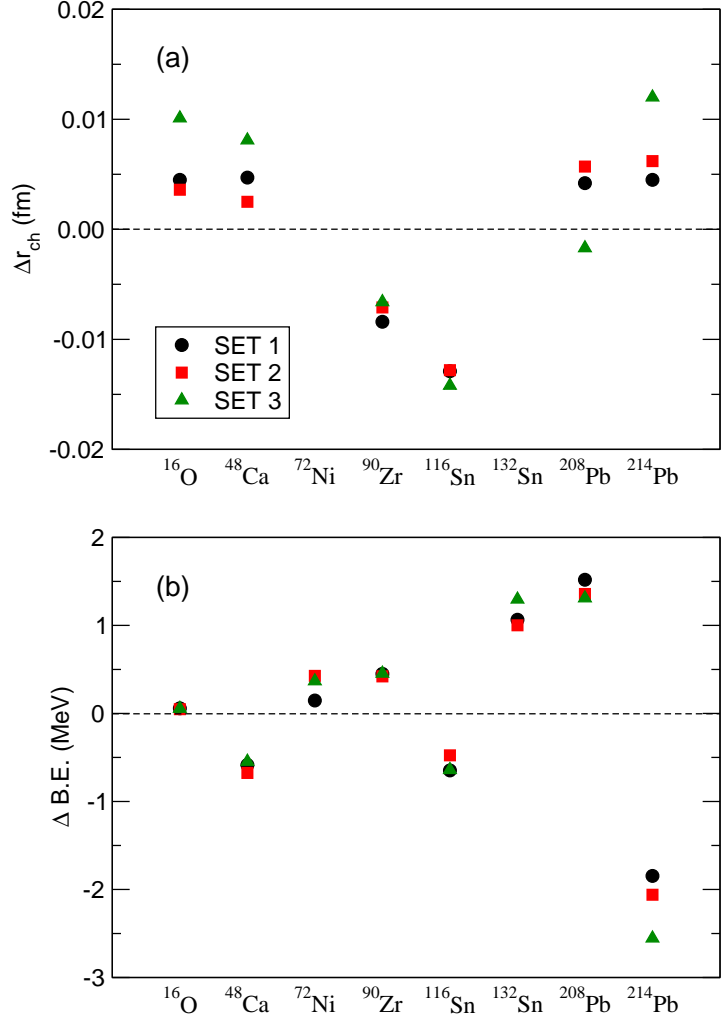} 
\begin {center}
\caption{\label{fig:finite-nuclei} (Color online) Differences between the theoretical
and experimental  values for the charge radii (panel (a)) and binding energies 
(panel (b))  of the eight nuclei used to adjust the parameters of the functionals 
{\sc SET 1}, {\sc SET 2}, and {\sc SET3}.
}
\end{center}
\end{figure}

%----------------------------------------------------------------------------------------------------------
This effect is illustrated in Figs.~\ref{fig:EoS-SNM-PNM} and \ref{fig:finite-nuclei}, in which we display the results for 
nuclear matter and finite nuclei obtained with the functionals determined by the parameters {\sc SET 1} (ten parameters), 
{\sc SET 2} (nine parameters), and {\sc SET 3} (eight parameters), 
in comparison to the original functional DD-PC1 (ten parameters). 
Note that DD-PC1 was actually not adjusted to the nuclear matter EoS and/or  
spherical nuclei, but rather to the binding energies of 64 axially deformed nuclei
in the mass regions $A\approx 150-180$ and $A\approx 230-250$ \cite{NVR.08}.

%----------------------------------------------------------------------------------------------------------
%----------------------------------------------------------------------------------------------------------
In Fig.~\ref{fig:EoS-SNM-PNM} we plot the equation of state for symmetric nuclear matter
(panel (a)) and neutron matter (panel (b)). The curves calculated with DD-PC1 
and the three functionals adjusted in this study ({\sc SET 1}, {\sc SET 2}, and {\sc SET3}), 
are compared to the microscopic equations of state~\cite{APR.98}. All three functionals, adjusted to 
the points shown in the figure (with the adopted uncertainty of 10$\%$), reproduce the equations 
of state with comparable accuracy. It is only in the region of extrapolation at higher densities, where 
no fit points have been specified, that their predictions start to diverge. This is similar to the results 
obtained in Ref.~\cite{NV.16}. DD-PC1 was not adjusted to these equations of state and, of course, 
does not reproduce the microscopic EoS particularly well, especially the EoS of neutron matter. 
Empirical properties of symmetric nuclear matter at saturation (binding energy, density, incompressibility), 
as well as two points at low and high density, are built into the parameters of DD-PC1 and, thus, this 
functional reproduces the microscopic EoS up to and slightly above saturation density.   
Figure \ref{fig:finite-nuclei} displays the absolute differences between the theoretical values 
and data for the charge radii and binding energies 
of the eight nuclei used in the least-squares fits of the parameter sets 
{\sc SET 1}, {\sc SET 2}, and {\sc SET3}. The binding energies are reproduced by all three 
functionals with similar accuracy, whereas several charge radii ($^{16}$O, $^{48}$Ca, 
$^{208}$Pb and $^{214}$Pb) calculated with the functional {\sc SET3} are markedly different 
from those predicted by {\sc SET 1} and {\sc SET 2}. This already indicates that with a further 
reduction of the number of parameters it would not be possible to accurately reproduce the data set.

%============================================================
%  Section 5
\section{\label{sec-summary} Summary and outlook}

One of the most important current research topics in low-energy nuclear physics is the development of a universal 
energy density functional framework that can be used in global studies of structure phenomena in different regions of 
the nuclear mass table. Even though structure models based on EDFs can accurately reproduce a variety of 
measured nuclear properties and in many cases provide useful predictions for regions far from stability where few data are available, 
empirically it has been known for a long time that nuclear EDFs exhibit an exponential range of sensitivity to parameter variations, 
crucially depend on just a few parameter combinations, while the remaining 
combinations of bare model parameters can only approximately be constrained by available data. Various approaches to 
the construction of EDFs lead to different, and sometimes very complex, functional dependence on nucleonic densities 
and currents, characterized by a relatively large number of parameters whose values are difficult to accurately determine either 
microscopically or from experiment. Parameter uncertainties and propagation of errors, as well as correlations between parameters, 
have been the subject of numerous recent studies in the framework of nuclear density functional 
theory (see, for instance, Ref.~\cite{JPhysG.15} and references cited therein).

In Ref.~\cite{NV.16} and in this work, we have analyzed a representative semi-empirical relativistic EDF, 
with a microscopically motivated ansatz for the functional density dependence, and parameters determined by empirical properties of 
homogeneous nuclear matter and data on ground-state nuclear properties. If the space of model predictions is considered as 
a manifold embedded in the data space, with model parameters as coordinates of the manifold, geodesic paths can be used to 
explore the boundaries of the model manifold. These are defined by points on the manifold where the metric (Fisher information 
matrix) becomes singular. Starting from a best-fit point for a given model functional, obtained by minimizing the penalty function $\chi^2$ 
that provides a measure of the distance in the data space between model predictions and the data point to which the model is 
fitted, one can explore the boundaries of the model manifold in the directions of eigenvectors of the Hessian matrix of 
second derivatives of $\chi^2$ at minimum. In \cite{NV.16} we have shown that the widths of the 
manifold of predictions for the model functional, that is, the arc lengths of geodesics along the 
eigendirections of the Hessian matrix, exhibit an 
exponential distribution nearly identical to the exponential range of sensitivity of the model to parameter variations. This is 
characteristic of sloppy models, and indicates that our model functional could contain functionals of lower effective dimension in 
parameter space that can equally well reproduce data, and with parameters more tightly constrained by the data.   

The general problem of a reduction of a nonlinear sloppy model functional to lower dimension in parameter space is difficult and 
depends on the data that are used to determine both the functional form of the density dependence as well as the values of 
the parameters. The solution cannot be found by simply eliminating bare model parameters but necessitates, 
often nonlinear, transformations in parameter space that also modify the form of density dependence. 
We have shown that the recently introduced Manifold Boundary Approximation Method (MBAM) \cite{Transtrum.14} 
can be used to systematically reduce the complexity and the sloppiness of a general nuclear EDF, with reduced 
sets of parameters constrained by the underlying microscopic dynamics and fine-tuned to nuclear data. While in 
Ref.~\cite{NV.16} we only considered a set of pseudo-observables that characterize the microscopic EoS of nuclear 
matter, in the present study the data set has been extended with points on the microscopic EoS of neutron matter and 
ground-state properties 
(binding energies, charge radii, difference between radii of neutron and proton distributions) of eight spherical nuclei. 
Since first and second derivatives of observables with respect to parameters along geodesic paths on the model manifold 
have to be computed, the inclusion of data on finite nuclei makes the application of the MBAM computationally much more challenging. 
Nevertheless, it has been possible to reduce our original ten-parameter functional to an eight-parameter functional that 
reproduces the given data set with comparable accuracy. An important result is that the functional density dependence of the 
coupling functions (density-dependent parameters) has been simplified to a polynomial form in the isoscalar channel of the 
functional. After two MBAM iterations, in the third the algorithm could no longer identify the boundary of 
the model manifold in the direction of the softest eigenvector of the Hessian matrix and, therefore, the dimension of 
the parameter space could not be reduced further. This is one MBAM iteration less than in our previous study \cite{NV.16},  
in which only pseudo-data on nuclear matter were used in the nonlinear least-squares fit. Obviously the additional data on 
finite nuclei place more stringent constraints on the functional form and parameter values, especially in the isoscalar channel, 
and thus prevent further parameter reduction.   

Even though we have only analyzed a single representative example 
of semi-empirical functionals currently used in nuclear structure studies, the illustrative study has shown how the MBAM 
can be used in the development and optimization of nuclear EDFs. In particular, this method can be applied to fully microscopic 
functionals that encode the underlying many-body dynamics in a complex dependence on nucleonic densities and currents, 
and include all terms allowed by symmetries. Generally such a functional will be characterized by a rather large number 
of parameters whose values have to be determined by low-energy data and, therefore, one expects that a microscopic 
derivation will produce a sloppy functional. When the complexity and parameter  
space of a general sloppy functional is systematically reduced by applying the MBAM as described in this work, it is 
the data that such a model is designed to reproduce that determine, not 
only the values of model parameters, but also the optimal functional form of the density dependence.  
%----------------------------------------------------------------------------------------------------------------
\begin{acknowledgements}
This work has been supported in part by 
the Croatian Science Foundation -- project ``Structure and Dynamics
of Exotic Femtosystems" (IP-2014-09-9159) and by the QuantiXLie Centre of Excellence.
\end{acknowledgements}
%--------------------------------------------------------------------------------------------------------------------------------------
%========================================================================================================

%--------------------------------------------------------------------------------------

\end{document}